# On the Possibilities of Hypercomputing Supertasks[1]

Vincent C. Müller
Anatolia College/ACT
www.typos.de

December, 2010

**Abstract**

This paper investigates the view that digital hypercomputing is a good reason for rejection or re-interpretation of the Church-Turing thesis. After suggestion that such re-interpretation is historically problematic and often involves attack on a straw man (the 'maximality thesis'), it discusses proposals for digital hypercomputing with "Zeno-machines", i.e. computing machines that compute an infinite number of computing steps in finite time, thus performing supertasks. It argues that effective computing with Zeno-machines falls into a dilemma: either they are specified such that they do not have output states, or they are specified such that they do have output states, but involve contradiction. Repairs though non-effective methods or special rules for semi-decidable problems are sought, but not found. The paper concludes that hypercomputing supertasks are impossible in the actual world and thus no reason for rejection of the Church-Turing thesis in its traditional interpretation.

---

[1] I am very grateful to Paul Benacerraf, Adam Elga and Athanssios Kehagias for illuminating discussion. I am also very grateful to two anonymous reviewers for *Minds and Machines*, and two for the *British Journal for the Philosophy of Science*.

## 1. Introduction: Church-Turing and Hypercomputing

### 1.1. Copeland and the Church-Turing Thesis

The philosophical literature on the notion of computing, whether it is in the context of computationalism in the philosophy of mind, the possibility of artificial intelligence or of computing machines in general, has traditionally assumed as background consensus that what a computer can do in principle is identical to what is "effectively computable", i.e. what can be computed by the mechanical application of a definite rule of finitely many instructions – of an algorithm.[2] The notion of computability was accordingly defined by Church, Turing and others in what is now known as the "Church-Turing thesis", one formulation of which is: *all and only the effectively computable functions can be computed by a Turing machine*. Strictly speaking, Church's thesis is that all effectively computable functions are recursive, and Turing's thesis is that all effectively computable functions are computable by the Turing-machine. Since the inversions to both theses are known to be true, to call a procedure "effective", "algorithmic", "recursive" or "Turing machine computable" all comes down to the same.[3]

In a series of papers, Jack Copeland and others have said that this traditional interpretation of the Church-Turing thesis is a misunderstanding, arguing that the Church-Turing thesis says nothing about what is computable *by machines*, or computable *in principle*, but it concerns only what can be computed *by humans*. As we shall see presently, this re-interpretation is motivated by the notion that machines, unlike humans, are capable of 'hypercomputing' and the Church-Turing thesis must thus be re-interpreted in order not to come out false. All sorts of errors in the philosophy of computing and mind are blamed on this alleged misunderstanding of the thesis (Copeland, 1997; 1998; 2000; 2002a; 2002b; 2003; 2004; Copeland and Proudfoot, 1999; 2000; cf. Shagrir and Pitowsky, 2003). If this interpretation were correct, one would have to distinguish one notion of computing for both humans and machines ("effective", "Turing machine computable") and a wider one for machines only ("algorithmic", "recursive").

---

[2] Indicative for the philosophy of mind: (Churchland, 2005; Fodor, 2000; Piccinini, 2004; 2007; Pinker, 2005; Scheutz, 2002); for artificial intelligence: (Copeland, 1993); for mathematical logic: (Boolos, et al., 2007, ch. 3ff).

[3] (Church, 1936; Turing, 1936; cf. Boolos, et al., 2007, ch. 3ff; Harel, 2000). The notion of "Turing machine" is well explained in many places, see particularly (Penrose, 1989, ch. 2; Floridi, 1999, p. 26ff; Davies, 2000, ch. 7; Copeland, 2003, p. 4ff).

### 1.2. Church-Turing and Church and Turing

Concerning the historical question what Church, Turing and other contemporaries had in mind, Copeland rightly points out that in the 1930ies and 40ies, the word "computer" meant a person doing computation, which is an indication that Copeland's historical thesis might be correct. While it is true that universal computing machines did not exist before 1941 (the "Z3"), there had been non-electronic calculating machines for centuries and Turing, of all people, was surely aware of the possibility of programmable (universal) computing machines. There are strong indications that Church (1936) and Turing thought the thesis to apply to machines, too – the main motivation for saying otherwise appears to be not to have Church and Turing say something wrong.

Turing states in the opening paragraph of his famous paper "'On computable numbers …": "The 'computable' numbers may be described briefly as the real numbers whose expressions as a decimal are calculable by finite means. … According to my definition, a number is computable if its decimal can be written down *by a machine*." [my emphasis] (Turing, 1936). About this paper, he said in a 1947 address to the London Mathematical Society: "I considered a type of machine which had a central mechanism, and an infinite memory which was contained on an infinite tape… One of my conclusions was that the idea of a 'rule of thumb' process and a 'machine process' were synonymous." (Turing, 1992, p. 106) (see also Hodges, 2006). His emphasis is on the 'mechanical' nature of the process, not on who or what carries it out.

Finally, it would appear to be precisely the point of Turing's 1936 paper to show that all effectively computable functions are computable by his machine, and thus that the halting problem of his machine is the *Entscheidungsproblem*. So it would be odd to have the *Entscheidungsproblem* for humans, but not for machines. (It would also constitute a dramatic inversion of the Penrose/Lucas argument, which says that machines have the *Entscheidungsproblem*, but humans do not.)

**1.3. Church-Turing and the "Maximality Thesis"**

Copeland focuses on the Church-Turing thesis for machines and calls this part of the traditional strong interpretation the "maximality thesis", stating it as follows: "all functions that can be generated by machines (working on finite input in accordance with a finite program of instructions) are Turing machine computable" (2000, p. 15). He says that while the Church-Turing thesis is true of humans, the maximality thesis is "known to be false" if we take the machines to be "machines in a possible world" (Copeland, 2000, pp. 15, cf 31). "It is straightforward to describe abstract machines that generate functions that cannot be generated by the UTM [Universal Turing Machine]" (Copeland, 2004, p. 12). What remains contentious on his view is merely whether the maximality thesis is true in the actual world.

Before we enter into the details, we must specify two senses in which this may well be correct – but which are beside the point, in my opinion.

First, we know that the set of all functions (even of all functions over the positive integers) is larger than the set of Turing-computable functions, since the former is not denumerable, while the latter is. But our issue here is not to delineate certain classes of functions. What we want to find out is whether it is indeed "straightforward to describe abstract machines" that *compute* such functions; that is to describe systems whose states are causally determined by their previous states (only then do they deserve the name of machines). If not, the burden of proof would be shifted onto those who want to reject the traditional strong reading of the Church-Turing thesis.

Second, the Church-Turing thesis concerns only *digital* or "discrete state" computing. This follows directly from the restriction to effective algorithmic procedures, which proceed "step by step", where steps are distinguished by a discrete state. This is not to say that the inversion holds: one might well hold that some digital procedures are not effective; for example those of hypercomputing. Whether non-digital, i.e. "analogue", or "continuous" computing deserves the name of "computing" and whether analogue mechanisms could compute functions that are not Turing-computable are matters not relevant to our point here – but see (Müller, 2008) for a discussion. As Hava Siegelmann (Siegelmann, 1995; 1997; Siegelmann and Sontag, 1994) and others have shown, there is good reason to believe that analogue mechanisms are possible which can compute functions that are not Turing-computable.

Both of these points would not refute the traditional strong interpretation of the Church-Turing thesis, however. The situation would still be very aptly described by Floridi, when he says: "From Turing power up, computations are no longer describable by algorithms" (1999, p. 36).

Accordingly, the strong Church-Turing thesis under discussion here is not identical to Copeland's "maximality thesis", since that thesis is restricted to machines. What is more, the strong Church-Turing thesis does not even imply the "maximality thesis", since the latter makes no mention of algorithms – an absence that is used by Copeland to attack it with the possibility of analogue computers. The "maximality thesis" is a straw man, and it is false.

### 1.4. Hypercomputing

The rejection of the Church-Turing thesis under its strong interpretation is motivated by the idea that there could be machines that could compute what no human and no Turing machine could compute, and this computing of what is not Turing-machine computable is now called "hypercomputing".[4] Proposed designs for machines include Turing's "O-machines" ("oracle machines" with a black box that answers non-computable queries non-mechanically[5]), "Zeno machines" (that can compute infinitely many steps, see below), analogue computers (but see above), quantum computers, Putnam-Gold machines (computers that can "change their mind"), probabilistic machines, machines in Malament-Hogarth universes, machines using the expansion of "mixmaster" universes and others. Despite all these proposals, it is probably fair to say that the various defenders of hypercomputing have not themselves proposed a notion of computing, they have restricted themselves to a rejection of the notion of computing expressed in the strong Church-Turing thesis.[6]

---

[4] Note that it is strictly speaking misleading to talk about the computing of a "Turing machine" in this context. A Turing machine is a theoretical device that can perform a particular algorithm and the theoretical universal Turing machine is a theoretical machine that can perform whatever any particular Turing machine can perform, i.e. it can be programmed to perform any algorithm. The Church-Turing thesis concerns the possibilities of this universal Turing machine and its relation to the notion of "effective computability". However this machine is just a *model* for what any mathematician with enough time and resources (paper and pencils - or tape and a read/write device) on his/her hand can compute. So, while the computer on my desk is a universal computer, its abilities are the same as that of the universal Turing machine (save its limited memory), but it is misleading to shorten this property to "it is a Turing machine".

[5] O-machines are mentioned by Copeland, but they just serve a theoretical purpose in Turing, they are not a proposed design for a computing machine. For a discussion, see (Cotogno, 2009).

### 1.5. Possibilities of Hypercomputing

The discussion about hypercomputing has focused on the question whether hypercomputing is possible in the actual world, given the physics of this world. A negative answer is sometimes called the "physical Church-Turing thesis" (e.g. Cotogno, 2003) or also, "Gandy's thesis" (after Gandy, 1980). There are many interesting problems with the view that such hypercomputing machines are possible in our world, given that the extant proposals involve infinity, such as infinite memory, or infinitely large machines, infinitely many steps, infinitely small parts, infinitely fast movement, infinitely fast information transfer, infinite amount of information transfer, infinitely precise measurement of quantum states, survival of infinite-energy states, infinitely expanding universes, etc.[7]

However, as long as no particular proposal is accepted this discussion can make no headway on the general question of whether hypercomputing is possible. After all, even if one rejects a particular proposal, it is prudent to remain agnostic about the possibility of a more ingenious design. While that discussion is going on, one has to accept that it is important to distinguish between the truth of the strong (traditional) and of the weak (Copeland's) interpretation of the Church-Turing thesis, since one is discussing whether a particular proposal falls under the one but not under the other. In order to secure the traditional strong reading of Church-Turing, one would have to show that hypercomputing is impossible in the actual world, or even in any possible world. Some attempts to refute the physical Church-Turing thesis been made (esp. Cotogno, 2003), using Cantor's diagonal technique, but these have been rebutted successfully (Ord and Kieu, 2005; Welch, 2004), in my opinion. I will make a new attempt to shift the burden of proof onto the supporters of infinite hypercomputing.

---

[6] Very useful surveys are in (Copeland, 1997; 2002b), more critically (Cotogno, 2003), also (Potgieter, 2006) for the more mathematical literature. Special issues in *Minds and Machines* 12 (2002) and *Theoretical Computer Science* 317 (2004).

[7] (Barrow, 2005, ch. 10) has a useful basic survey. For a quantum proposal, see (Kieu, 2002; Ord and Kieu, 2005). For a relativistic proposal, (Shagrir and Pitowsky, 2003), cf. also (Potgieter, 2006). For a proposal of "shrinking" Zeno-machines in a Newtonian universe, see (Davies, 2001).

My impression from the mathematical literature is that there is little hope to prove hypercomputing contradictory and thus impossible in any possible world. – But on the other hand, the distinction between logical and actual possibility might not be so clear after all: "The misconception is that the set of computable functions (or the set of quantum-computational tasks) has some a priori privileged status within mathematics. But it does not. The only thing that privileges that set of operations is that it is instantiated in the computationally universal laws of physics. It is only through our knowledge of physics that we know of the distinction between computable and non-computable […], or between simple and complex." (Deutsch, 2004, p. 99)

## 2. Zeno Machines: Infinite Hypercomputing

Let us investigate the notion of a "Zeno machine", a concept proposed by Hermann Weyl (1927). A Zeno machine is specified in such a way that each step takes a fraction of the time of its predecessor, so if the first step takes ½ a second, for example, the times for each step could be: ½, ¼, $1/8$, … This machine could make a denumerable infinity of computing steps in finite time, in one second. It starts at time $t_0$, then runs through a series of steps $t_n$ and is done at time $t_1$. This machine shows clearly that we need to distinguish "in finitely many steps" from "in finite time" in the formulation of the Church-Turing thesis.

Zeno machines are repeatedly presented by Copeland as examples of possible hypercomputers (called "accelerating Turing-machines"), and they are the most intensely discussed proposal for digital hypercomputing (cf. Ord and Kieu, 2005). Zeno machines are not standard Turing machines since the latter produce results only once they halt, after a last step (though they can be set to motion again, even infinitely many times), while Zeno machines can go through infinitely many steps – though they will be "done" in a different sense, namely *in time*.

## 2.1. Background: Supertasks

The logical possibility of a physical object carrying out infinitely many tasks (e.g. computing steps) in finite time was much discussed in the 1950ies and 60ies in the context of Zeno's paradoxes of movement (esp. Achilles and the tortoise, and the racetrack) and such tasks were dubbed "supertasks" by James Thomson (1954). In order to show that performing supertasks is impossible, Thomson had proposed to consider a lamp that is switched on and off infinitely many times. He then said that from the assumption that each time the lamp is switched on it is also switched off afterwards, it follows that it can be neither on nor off after the switchings are over - which he claimed to be a contradiction. Paul Benacerraf (1962, p. 779ff) criticized this move, pointing out that, given the specification, nothing follows from the states of the lamp inside the series about the state of the lamp after the series. This criticism is widely regarded as correct.

The logical gap between what is the case inside the infinite series and what is the case after the series is crucial for the following discussion and I shall call it the "Benacerraf gap". I propose that the defender of infinite hypercomputing has to bridge the Benacerraf gap, in order to generate an output – and that is the problem which is ignored. The description of a Zeno machine is indeed unproblematic, but as soon as it includes a device that 'bridges the gap', contradiction looms, as we shall see presently.

It is crucial for the understanding of the Benacerraf gap to keep in mind that there is no such thing as "the last step" or "the last state" in the series, and accordingly, no last step that can determine the state of the lamp. Also, for any point in time arbitrarily close to time $t_1$, there is still a further step to take place later. Given that there is no last state, one cannot measure/read out the last state and one can not write a program that instructs "do the last step and then do this and halt", neither can we ask "what is the state after the last step?"

So, a first form of the fundamental problem is that a) we cannot have a computational output after the "last step", but b) neither can we just look at the output after the series is over in time, since "nothing follows", as Benacerraf had pointed out. So, whatever the state of the Zeno machine at $t_1$, how can it be the *effect* of the infinite t-series? Can we make sure that there is an output that can be generated without reliance on contradictory notions like "the last step in an infinite series"?

As Benacerraf says, "Certainly, the lamp must be on or off at $t_1$ (provided it has not gone up in a metaphysical puff of smoke in the interval), but nothing we are told implies what it is to be." (Benacerraf, 1962, p. 768).

I will suggest that we are faced with a dilemma: either we have a machine where "nothing follows", or we have a machine that bridges the Benacerraf gap but computes impossible results. Let us first set up the situation with machines that can bridge the Benacerraf gap.

**2.2. A Proposal for Infinite Computing: Facing the Benacerraf Gap**

One might, for example, want to know the answer to Brouwer's classic question (discussed repeatedly by Wittgenstein) whether there is a sequence of "777" somewhere in the infinite expansion of π. This problem cannot be computed by a Turing machine because a negative answer would require looking at all of the infinite expansion of π. However, a positive answer *is* possible if one comes across the sequence "777" somewhere in π – in fact this has happened, and we now know that 777 does indeed occur in that expansion. Many famous mathematical problems have this "semi-decidable" feature, e.g. Hilbert's Tenth Problem (claimed to be solvable by probabilistic quantum computing in Kieu, 2002; 2004) and Turing's halting problem. Since the halting problem is precisely the problem whether the Turing machine will halt on a given problem, the *Entscheidungsproblem* itself is one of these problems.

Copeland seems to think that a semi-decidable task is computable by Zeno machine in the following fashion: Our hypercomputer may be fitted with a lamp and, for example, programmed in such a way that it switches on the lamp as soon as it finds the sequence "777" in π. After the series of computing is over, at $t_1$ or later, you look at the lamp: if it is on, there is such a sequence, otherwise there is not. In this fashion, any Boolean (true/false) decision over infinite domains could be settled. (And it would appear that *any* formal problem that can be formulated in binary code could be settled.)

Recall, however, that nothing followed from the specification of Thomson's lamp about the state of his lamp at $t_1$ or later. Is this any different with our new, separate, indicator lamp? What the specification does tell me is that I can check whether the lamp is on at any time in the t-series, arbitrarily close to $t_1$: if the lamp is on, a "777" has been found. But *this* task, namely whether the sequence is to be found in π *up to a specific point*, is a Turing-computable task. Does the specification of our machine tell me what is the case with my lamp at $t_1$ or later? No, it does not. We have no reason to take the state of such a lamp as the output of the machine. More work needs to be done if we want to bridge the Benacerraf gap.

If one wanted to provide a specification that bridges the gap, one has to avoid *any* reference to a "step", and instead talk about what is the case "after the series is over in time". One way to achieve this is to include in the specification that there is an *indicator* (like the "lamp" above) separate from the actual machine, and to add a *bridging principle* to the effect that "the indicator is wholly determined by the machine", in particular, it does not change other than by action of our machine. We can then check the indicator (a variable to read out, a lamp, or a display) after $t_1$ and use this indicator for the output of computing results. This bridging principle does the job of what Earman and Norton call the "persistence property" of the natural world, the property of persisting unchanged after the t-series (Earman and Norton, 1996, p. 238ff). This property causes the apparent contradiction in Thomson's lamp, on their analysis – and it is the property that hypercomputing has to re-instate … calling for trouble.

**2.3. Beyond the Benacerraf Gap – Into the Abyss**

So, the bridged indicator might get us across the Benacerraf gap, but do we really want to go there? In his 1954 paper, Thomson had also proposed a machine that prints the digits of π on a tape which is generated at the same speed as the computation. After the end of the computing series, we would have an infinitely long tape with each digit of π printed on it. He additionally proposed a parity machine connected to the π-machine, and asks "what appears on the dial after the first machine has run through all the integers in the expansion of π?" (Thomson, 1954, p. 5) – pointing out that any output is contradictory. So, would bridging the gap not have the unacceptable consequences Thomson wanted to warn us about? It appears that we would now be able to compute impossible things like the highest natural number, the parity of the last digit of π, the result of "0+1-1+1-1…", etc. Copeland concedes that this combination with a parity machine is logically impossible, and also concedes: "… Thomson's query as to what state an infinity machine may consistently be supposed to be in *after* it completes its supertask is a good one." (Copeland, 2002a, p. 286f.). Indeed it is – in fact it is crucial. Copeland then uses what I call the Benacerraf gap and says: "The answer to the Thomsonian question 'Where is the scanner at that point?' is: Nowhere." (Copeland, 2002a, p. 289). But that is not an option. We are told is that this machine is computing, but that we can not have an output, necessarily! I think it will be agreed that a machine that necessarily has no output does not qualify as a computing machine: it is "hypocomputing" rather than hypercomputing. The dilemma is that if we do not combine the first machine with the 'second' machine, we do not bridge the Benacerraf gap and we do not get an output, but if we *do* bridge, contradiction looms.

Let us illustrate the second horn of the dilemma by a closer look at a bridged π-machine:

1) A Zeno machine will compute an infinite
   number of steps in finite time                                    (Assumption)

2) There is a program (P) such that:
   a) it calculates the digits of π one by one, and
   b) it writes each calculated digit into a variable (N), and
   c) (N) changes if and only if (P) changes it                      (Assumption)

3) A Zeno machine can run (P)                                        (Assumption)

4) After carrying out (P) on the Zeno machine,
   variable (N) holds one digit of π (D)                             (from 2 and 3)

But which digit is that (D)? After (N) has completed its own supertask, (D) cannot really be any particular digit out of the infinite expansion of π. One way to put this problem is this:

5) (D) is the last digit computed in time
   or
   (D) is not the last digit computed in time                        (from 4, Assumption)

But now we can see, that either of these options is unacceptable:

6) If (D) is the last digit computed in time,
   then it is the last digit of π                                    (from 1, 2 and 4)

7) There is no last digit of π                                       (Assumption)

8) (D) is not the last digit computed in time                        (from 6 and 7)

9) If (D) is not the last digit computed in time, then there is
   a digit computed later.                                           (from 1, 2 and 4)

10) There is no digit computed later than (D)                        (from 2 and 3)

11) (D) is the last digit computed in time                           (from 9 and 10)

11) contradicts 8) and thus 5) must be false, and so must 4) – as we indicated already. In order to get out of this problem, one might drop any part of the assumptions 1), 2), or 3); say the bridging principle in 2c) is insufficient; or say 4) does not follow. (If there is a worry about 4) being internal, you may add a line to the pro-gram (P) where a further variable (M) is set to the value of (N) and then read out (M) after the series.) All of these moves, however, remove the necessary bridging and result in a machine with no output.

More importantly, one might wish to drop 2a) and deny that the Zeno machine proceeds step-by-step, i.e. that it computes 'effectively' in this sense. This is a possibility but it removes the motivation to infer anything about Church-Turing from such non-effective Zeno machines. The argument would still show that effective supertask hypercomputing can run into contradictions. If one wanted to use non-effective methods, the question arises how 'bridging' can be achieved, i.e. how the state of the output indicator can be considered as caused by the Zeno-machine.

To put the argument in terms of a lesson from Thomson vs. Benacerraf: Thomson argued that supertasks would result in impossible states, and Benacerraf showed that nothing about the resulting states followed from the specification of Thomson's machines (e.g. his lamp that is switched on and off), while he conceded that these states are indeed impossible. To get a Zeno machine to compute, we must specify it such that something *does* follow about the resulting states (so that we can take them as output), but then we are back at Thomson's impossible states.

So, while Copeland could say "No inconsistency in the notion of a π-machine was ever demonstrated" (2002a, p. 284), we now have a dilemma of computing with no output or bridged computing with a contradiction.

Perhaps this problem even generalizes to all other hypercomputers and the question arises whether their feat of "completing" infinite steps in finite observer-time (through fractions of time, quantum superpositions, relativistic space-times, or whatever) does not allow for the same paradoxical results. (For example, any Zeno hypercomputing machine that counts its own infinite steps would have to calculate the "highest natural number".)

## 3.  Objections to the Proposed Dilemma

### 3.1. Infinity vs. Hypercomputing in Mechanisms

It may be thought that this second horn of the hypercomputing-dilemma must show too much: could an infinite omniscient God not know mathematical facts over the infinite, are there no functions with truth values over the infinite? Indeed there are, as we granted earlier, but that is not at issue. For example, there is nothing inconsistent in the notion of a God digitally going through the extension of π in a minute. In particular, if one removes the condition of the computational output (as does Shagrir, 2004, p. 110f), no contradiction ensues. There is also nothing impossible about a logical consequence from God going through some infinite steps. But what even God cannot do is to perform an infinite effective *digital* computation, say, hold up one of ten fingers (the original digit "indicator") each time he computes a digit of π *and* claim that nothing else changes the state of his fingers after the computation (a bridging principle). What would his fingers show once God is done the extension of π? We have seen that an infinite computing machine with a bridging principle is impossible – at least if that machine writes each digit of π, or keeps a counter, or switches an indicator lamp on/off after every +1/-1 computation, etc.

The possibility of hypercomputing involves more than a formal specification of the algorithm that is free from contradiction; it involves the possibility *as a digital computing mechanism,*[8] i.e. as a mechanism in which the state of the *output* (e.g. indicator at $t_1$ and after) is *causally determined* by the step-by-step workings of the mechanism. Put in these causal terms, a bridged supertask is one where the supertask has an effect that lasts beyond the time of the completion of the task, an effect that can be taken as the output of the effective computation. Since this argument involves causation it concerns the physical possibility of such machines, not the logical possibility of their specification.

The burden of proof is now on the defenders of hypercomputing, who have to explain the specification of the machine such that it does have an output but does not result in contradiction. They must explain how such an output can be the result of the total digital computation without being the result of effective step-by-step computation and while avoiding the programs for a Zeno machine that result in contradiction.

---

[8] I use "machine" and "mechanism" interchangeably in this paper, for lack of an adjective in English that differentiates the property of a mechanism ("mechanical") from that of a machine ("machinical").

### 3.2. Hypercomputing Semi-decidable Problems Only?

For a machine that deals with semi-decidable problems, there is hope for specifying such a reason, since the results themselves cannot be a source of contradiction: no single answer to a well-formed true/false question results in a contradiction. The only difference between the manifestly impossible machines and the proposed machines for semi-decidable tasks appears to be that the bridged indicator is changed infinitely many times in the former and only once (if at all) in the latter - though perhaps infinitely close to $t_1$. In fact, the indicator itself is performing a supertask in the impossible machines. Perhaps the defenders of hypercomputing could come up with a principle that would allow for the semi-decidable machines, but rule out the manifestly impossible machines? Let us have a brief look at some instructive candidates:

a) Nothing can be the effect of infinitely many causes[9]

This would prevent the existence of a bridged indicator for the impossible machines that has been updated infinitely many times. It also implies that the Zeno-machine itself must go up in Bencerraf's "puff of metaphysical smoke" after the t-series, since its state at $t_1$ would be a result of infinitely many steps. More importantly, it would also show that the state of the indicator cannot be the effect of all of the infinite steps of the Zeno-machine. But the specification of the mechanism must be such that the state of the indicator lamp at $t_1$ can be taken as the "output" of the computing procedure. This applies when the lamp is "off" as well as when it is "on". *Not changing* the indicator after a particular computation must also be an effect if we want to take it as output – an effect of all the infinitely many computing steps. We cannot take the indicator as output if the causal connection was cut somewhere during the t-series and this cut caused the lamp to be "off". (Remember, we are talking about semi-decidable tasks, where the result "on" is computable in finitely many steps, only the result "off" is not.) This would return to the first horn of the dilemma: computing without an output. Therefore infinite hypercomputing would impossible. In fact, it implies a rejection of assumption 3) above, so it is not a way out.

---

[9] Meaning "… within finite time". There can be no effect after an infinite time (for an observer) anyway. (But see (Hamkins and Lewis, 2000) for an investigation of what is mathematically possible.)

b) Nothing can have infinitely many effects

This would prevent the Zeno-machine from updating the indicator infinitely many times. But it would also prevent any Zeno-machine from running, since starting it would have infinitely many effects. Caused supertasks and thus infinite hypercomputing would be impossible. This also implies a rejection of 3), so it is not a way out either.

c) If something changes infinitely many times, then it must go out of existence afterwards, without any effect

This would force the Zeno-machine itself to go out of existence after its activity (as is the case in the non-Newtonian proposals). It is weaker than a), but strong enough to prevent a bridged indicator that has been updated infinitely many times (resulting in contradictions), while possibly allowing a bridged indicator that is changed once. However, as noted under a) above, even a bridged indicator that is changed only once (or not at all) should be seen as being the result of the entire workings of the Zeno machine (probably even of all its infinitely many steps) – so the state of indicator for a semi-decidable task cannot be seen as the result of the Zeno machine after computation is over if c) were true.

Another issue is what causes this "going out of existence". If cannot be the consequence of internally "having completed" the supertask, but neither can it be caused from the outside by "time is up". On the other hand, if going out of existence is due to some gradual process, it becomes implausible that an object should have effects infinitely close to its going out of existence. – I think that explanation c) looks ad hoc, but we should be open to arguments sustaining it.

To sum up, bridged supertasks require that something can be the effect of infinitely many causes (~a), that something can have infinitely many effects (~b), and that the bridged supertask does have effects (~c). Denying one of the three conditions amounts to denying the possibility of bridged supertasks, thus the denial of 3) in the argument above - for any program (P). In other words, all three attempts at repair throw the baby out with the bath water: they disallow contradictory machines but also all other machines for semi-decidable problems. I do not see any further plausible way out. If there really is no other, we have to conclude that bridged supertasks are impossible - whether or not they are considered computing machines. To put it the other way around, if bridged supertasks were possible, infinite hypercomputing should have been possible. But infinite hypercomputing is not possible, so bridged supertasks must be impossible.

## 4. Conclusion: Zeno's Supertasks and Computing the Incomputable

If the general conclusion could be established that bridged supertasks are impossible (by a thorough rejection of all ways out), this would have ramifications for Zeno's classic paradoxes. If principle a) is true, it cannot be the case that moving through a stretch from a point A to a point B is to perform a supertask, since the arrival at B presumably is the effect of moving through that stretch - and we do not vanish as soon as we complete a movement from A to B. Equally, if we prefer only principle b), and take moving through a stretch from a point A to a point B as a supertask, then our movement cannot have a cause – which seems false. If c) is true, after all, we would have to vanish after a movement. So, on *any* of these three explanations Zeno must be wrong when he says that one movement is to make infinitely many movements. If any of these or any other explanations are proposed, such ramifications must be considered.

Concerning our original question, I conclude that Thomson was right that if anything follows from states inside the series to states outside the series, then contradiction ensues. And Benacerraf was right that nothing does follow from states inside the series to states outside the series – unless one adds a bridging principle. In other words, either the Zeno machine can be specified, does bridge the gap, but then its specification involves contradictions; or it is underspecified, does not bridge the gap, but then it does not compute an output. Either way, Zeno machine hypercomputers are impossible – and so are probably all bridged supertasks. Therefore, the notion of infinite digital hypercomputing is no reason to reject the traditional interpretation of the Church-Turing thesis.